\documentclass{elsart}
\usepackage{epsfig}
\usepackage{rotate}
\usepackage{amssymb}
\usepackage{amsmath}

\begin{document}

\begin{frontmatter}
\begin{center}
{\large \bf Invisible decay of orthopositronium vs extra dimensions} 
\end{center}
\vspace{0.5cm}

\begin{center}  
S.N.~Gninenko$^a$,\footnote{Sergei.Gninenko\char 64 cern.ch}
N.V.~Krasnikov$^a$,\footnote{Nikolai.Krasnikov\char 64 cern.ch}
A. Rubbia$^b$\footnote{Andre.Rubbia\char 64 cern.ch}
\\
\vskip0.5cm
{\it $^a$Institute for Nuclear Research of the Russian Academy of Sciences,\\ 
Moscow 117312}\\
{\it $^b$Institut f\"ur Teilchenphysik, ETHZ, CH-8093 Z\"urich, Switzerland}
\end{center}                                                            
 
\begin{abstract} 
In models of Randall-Sundrum type orthopositronium ($o-Ps$)
can disappear due to tunnelling 
into additional dimension(s). The experimental signature of 
this effect is the  invisible decay
of orthopositronium. We point out that this process  may  occur at a rate 
within two or three orders of magnitude of the present experimental upper 
limit. We discuss this in details and stress that the existence of invisible 
decay of orthopositronium in vacuum could
explain the $o-Ps$ decay rate puzzle. Thus,  our result 
enhances the existing motivation and justifies efforts for a 
more sensitive search for  $o-Ps\to invisible$ decay in a near future 
experiment. Possible manifestations of new physics in other
rare (exotic) decays of orthopositronium are also discussed. 
\end{abstract}
\end{frontmatter}
 
\section{Introduction}
Positronium ($Ps$), the positron-electron bound state,
 is the lightest known atom, which is bounded and self-annihilates 
through the same, electromagnetic interaction. At the 
current levels of experimental and theoretical precision this is  
the only interaction present in this system, see e.g. \cite{rich}. 
This feature has made positronium an ideal system for  
testing of the QED calculations accuracy
for bound states, in particular for the triplet ($1^3S_1$)
state of $Ps$, orthopositronium ($o-Ps$). 
Due to the odd-parity under
C-transformation  $o-Ps$ decays
predominantly into three photons. 
 As compared with singlet ($1^1S_0$) state (parapositronium),
 the "slowness" of $o-Ps$ decay rate, due to the phase-space and  additional
$\alpha$ suppression factors, gives an enhancement factor $\simeq 10^3$,
making it more sensitive to an admixture of 
new interactions which are not accommodated in the Standard Model.
 
The study of the $o-Ps$ system has a long history \cite{rich}.  However,
in spite of the substantial efforts devoted to the
theoretical and experimental determination of the $o-Ps$ properties,  
there is a long-standing  puzzle: the $o-Ps$ decay rate in vacuum 
measured by Ann Arbor group, 
$\Gamma^{exp}=7.0482\pm 0.0016 ~\mu s^{-1}$ \cite{nico}, has a 
$\simeq 5\sigma$ discrepancy with the  
predicted value $\Gamma =7.03830\pm 0.00007 ~\mu s^{-1}$
\cite{adkins1} (see also \cite{lepage}). 
This discrepancy has been recently confirmed by more precise calculations
of Adkins et al. \cite{adkins2}, 
 including corrections of the order $\alpha^2$.

 The results of the 
recent Tokyo measurements of $o-Ps$ decay rate
in low density SiO$_2$ powder corrected for matter effects \cite{asai}
agree, however within the errors  with
the result of \cite{adkins2}, thus making the situation confusing. 
So, it is difficult to disagree with the Adkins 
{\em et al.} \cite{adkins2} statement that : {\em...no conclusions can be 
drawn until the experimental situation is clarified}.

 Originally it was thought that an exotic $o-Ps$
decay, not taken into account in the calculation  of $\Gamma$
and given a relative contribution to the $o-Ps$ decay rate at the level 
of $(\Gamma^{exp}- \Gamma)/\Gamma \simeq 10^{-3}$ would solve the discrepancy.
At present, it is believed that practically all  {\em visible} exotic decays of
$o-Ps$ ( i.e. decays accompanied by at least one photon in the final state)
are excluded experimentally (for review, see e.g. \cite{scals,dobr}). However, 
there is still an intriguing explanation both of the discrepancy 
 and Tokyo results. The idea was first discussed by 
S. Glashow \cite{glashow} and is  based on  $o-Ps \to$ mirror o-Ps
 oscillations resulting in {\em invisible} decay of $o-Ps$ in 
vacuum, see Section 4 and ref. \cite{gnin,foot}.

 In this paper we point out that in  models with additional 
infinite dimension(s) of Randall-Sundrum type positronium can 
disappear into additional dimension(s). The experimental manifestation of 
this effect is the invisible decay
of positronium. Thus, this enhances motivation for further experimental
 search for this decay mode and makes it very  interesting and exciting.
   
The paper is organised as follows. In Section 2 we remind the 
results on o-Ps decay rate calculation in the Standard Model. In Section 3
we briefly review the experimental results and phenomenological 
models on exotic visible decays of o-Ps and improve some bounds.
In section 4 we discuss  phenomenological models for 
 $o-Ps\to invisible$ decays.  We  propose a model leading to
$o-Ps\to invisible$ decays which could be at the experimentally interesting 
level. We also point out that  models with infinite additional 
dimension(s) of Randall-Sundrum type predict the disappearance of 
orthopositronium as a result of tunnelling into 
additional  dimension(s). The estimate of 
corresponding transition rate is presented. 
Section 5 contains concluding remarks.

\section{Orthopositronium decays in Standard Model}

The three photon decay width of the orthopositronium is 
\cite{adkins1}-\cite{adkins2}, 
\begin{equation}
\begin{split}
\Gamma(o-Ps \rightarrow \gamma \gamma \gamma)& =
\frac{2(\pi^{2} -9)\alpha^6 m_e}{9\pi}[1 - 10.28661(1) \frac{\alpha}{\pi} 
-\frac{\alpha^3}{3} ln(\frac{1}{\alpha})  \\
 & + B_0(\frac{\alpha}{\pi})^2 
-\frac{3\alpha^3}{2\pi}ln^2\frac{1}{\alpha} + O(\alpha^3 ln (\alpha))]
\end{split}\end{equation}
The coefficient $B_0$ has been recently calculated to be 
$B_0 = 44.52(26)$  \cite{adkins2}, resulting in
\begin{equation}
\Gamma(o-Ps \to \gamma \gamma \gamma) = 7.039934\pm 0.000010 ~\mu s^{-1}
\end{equation}
The five photon $o-Ps\to 5 \gamma$ decay branching ratio is of order 
$\alpha^2$  \cite{lepa}
\begin{equation}
Br(o-Ps \to 5\gamma) = 0.19(1)(\frac{\alpha}{\pi})^2 
\approx 1.0\times 10^{-6}
\end{equation} 
and therefore does not significantly influence the total decay width.    
Experimental result on o-Ps $\to 5 \gamma$ decay gives \cite{mats} 
\begin{equation}
Br(o-Ps \rightarrow 5 \gamma) = [2.2^{+2.6}_{-1.6}(stat) \pm 0.5(syst.)]\times 10^{-6} 
\end{equation}
 in agreement with the theoretical prediction of Eq.(3).\\
Within the Standard Model orthopositronium can also decay weakly into
 neutrino-antineutrino pair.  
The $o-Ps \to \nu_e \bar{\nu}_e$ decay occurs through 
$W$ exchange in $t$ channel and $e^+e^-$ annihilation via $Z$.
The decay width is \cite{czar}
\begin{equation} 
\Gamma(o-Ps \to \nu_e  \bar{\nu}_e) = 
\frac{G^2_F \alpha^3m^2_e}{24\pi^2}(1 + 4 \sin^2\theta_W)^2 
\approx 6.2 \times10^{-18}\Gamma(o-Ps)
\end{equation}
For other neutrino flavours only $Z$-diagram 
contributes. For $ l \neq e$ the decay width is \cite{czar}
\begin{equation} 
\Gamma(o-Ps \rightarrow \nu_l \bar{\nu}_l) = 
\frac{G^2_F \alpha^3m^2_e}{24\pi^2}(1 - 4 \sin^2\theta_W)^2 
\approx 9.5 \times10^{-21}\Gamma(o-Ps)
\end{equation}
Thus, in the Standard Model the $o-Ps \to \nu \bar{\nu}$ decay width
is too small and its contribution to the total decay width can also 
be neglected.

\section{Visible exotic decays  of orthopositronium}  

Visible exotic  decays of $o-Ps$ can be classified into following
 categories: i)  $o-Ps\to \gamma  X$ ii)
 $o-Ps\to \gamma \gamma X$  iii) $o-Ps\rightarrow N \gamma$,
where  $X$ is a new light  
particle(s) and $N=2,4,..$ (for review, see e.g. \cite{scals,dobr}).


For the decay $o-Ps \to \gamma + ~X$,  where $X$ is a 
long-lived particle the experimental bound is \cite{sas} 
\begin{equation}
Br(o-Ps \rightarrow  \gamma +~X) < 1.1 \times 10^{-6},
(m_X < 800~keV)
\end{equation}
If $X$ decays within detector, e.g. into $ 2\gamma$, the 
experimental bounds depend on the $X$-particle mass \cite{tsu}-\cite{klubak}
\begin{equation}
Br(o-Ps \rightarrow \gamma +~X \to 3\gamma) 
< 2 \times 10^{-4}, (300~KeV < m_X < 900~keV)  
\end{equation}
\begin{equation}
Br(o-Ps \rightarrow \gamma +~X \to 3\gamma) 
< 2.8 \times 10^{-5}, ( m_X < 30~keV)  
\end{equation}
\begin{equation}
Br(o-Ps \rightarrow \gamma +~X \to 3\gamma)  
< 2 \times 10^{-5}, ( 900< m_X < 1013~keV)  
\end{equation}


Note  that all above experimental
 limits are based on a search for a peak in the energy 
spectrum of photons arising from the  2-body decay of $o-Ps$. In the case of
an exotic 3-body 
$o-Ps \rightarrow \gamma + X_1 + X_2$ decay (which still might be 
a solution of the $o-Ps$ decay puzzle \cite{scals,escri}), 
 the signal cannot manifest 
itself through the peak in the $\gamma$ energy spectrum,
thus making the limits for this decay mode weaker.
 The current
estimate for $Br(o-Ps\to \gamma X_1 X_2)$ from the results of the indirect experiment of ref. \cite{mits} is around $10^{-4}$ 
\cite{scals}.


\subsection{Exotic $o-Ps\to \gamma X$ decay}

Consider the model with light pseudoscalar \cite{kleymans} (e.g. axion)
which  predict the existence of the exotic decay 
$o-Ps \rightarrow \gamma +X$ decay.
The interaction Lagrangian  \cite{kleymans} is
\begin{equation}
L_X = g_X \bar{\psi}\gamma_5\psi X
\end{equation}
The  $o-Ps \rightarrow X \gamma$ decay width is determined by 
the formula \cite{kleymans}
\begin{equation}
\Gamma(o-Ps \rightarrow \gamma X) = \frac{8}{3}g^2_X \cdot 
\frac{\alpha^3 m^2_e}{8\pi}(1 - \frac{m^2_X}{m^2_{o-Ps}}) = 
g^2_X (1 - \frac{m^2_X}{m^2_{o-Ps}})\times 5.84 \cdot 10^4\mu s^{-1}
\end{equation}

Strong  bound on $\alpha_{Xe} \equiv g^2_X/4\pi $ arises from the results on 
anomalous magnetic moment of electron. 
The measurements of $a_e = (g_e -2)/2$  \cite{dyck} give
\begin{equation}
a^{exp}_{e^{-}} = 0.0011596521884(43),
\end{equation}
\begin{equation}
a^{exp}_{e^{+}} = 0.0011596521879(43)
\end{equation}
whereas the prediction is
\begin{equation}\begin{split}
a^{SM}_e& = \frac{\alpha}{2\pi} -0.32847844400(\frac{\alpha}{\pi})^2 
+ 1.181234017(\frac{\alpha}{\pi})^3 - 
1.5098(384)(\frac{\alpha}{\pi})^4 \\
& + 1.66(3) \times 10^{-12}
(hadronic~and~electroweak~contributions)
\end{split}
\end{equation}
The measurements (13,14) provide the best determination of the 
fine structure constant \cite{kinoshita}
\begin{equation}
\alpha^{-1}(a_e) = 137.035 999 58(52)
\end{equation}
Another determination of $\alpha$, from the quantum 
Hall effect (QHE)\cite{mohr}
\begin{equation}
\alpha^{-1}(QHE) = 137.036 003 00(270)
\end{equation}
has larger error. 
If the new light particle $X$ exists, it would  give
 contribution to the anomalous 
magnetic moment of the electron, but would not influence the determination of 
the fine structure constant from the QHE.   
The value of the fine structure constant extracted from the QHE leads to
\begin{equation}
a_e = .001 159 652  2172(229)
\end{equation}
The estimate of a possible additional contribution to the electron 
anomalous magnetic moment results in 
\begin{equation}
|a^{exp}_{e^-}-a (QHE)| \lesssim 4.6 \cdot 10^{-11}~~~~(90\% C.L.)
\end{equation}   

For $m^2_X \ll m^2_e$ the $X$-boson   contribution 
to anomalous magnetic moment of electron is 
\begin{equation}
(g - 2)_X \approx -\frac{1}{8\pi^2}(g^2_X)
\end{equation}
From the experimental bound (19) we find that $g^2_X/4\pi \lesssim 6\cdot 
10^{-10}$ and hence the bound is
$Br(o-Ps \rightarrow \gamma X) \lesssim 6 \times 10^{-5}$, 
that is  weaker than the current experimental bound (7).

\subsection{Exotic $o-Ps\to \gamma \gamma X$ decay}

Consider the case of $o-Ps \to \gamma \gamma X$ decay mode.
The negative search  for direct $e^{+}e^{-}$ annihilation 
$e^{+}e^{-} \rightarrow \gamma + X$ leads for 
the case of long-lived $X$ particle to the bounds \cite{mitsui}
\begin{equation}
Br(o-Ps \rightarrow \gamma \gamma  X) < 4.2 \times 10^{-6} ,
(m_X < 200~keV) 
\end{equation}
\begin{equation}
Br(o-Ps \rightarrow \gamma \gamma  X) < 2.1 \times 10^{-5} ,
(m_X < 2m_e) 
\end{equation}
In ref. \cite{skalzey} a more stringent bound 
\begin{equation}
Br(o-Ps \rightarrow \gamma \gamma X) < 4 \times 10^{-6}
\end{equation}
has been obtained.
Consider the model  with  vector $X$-boson and with 
the  interaction Lagrangian 
\begin{equation}
L_X =g_X \bar{\psi} \gamma_{\mu} \psi X^{\mu}
\end{equation}
One can find that for $m_X \ll m_e$
\begin{equation}
Br(o-Ps \rightarrow \gamma \gamma X) = \frac{3\alpha_{X}}{\alpha},
\end{equation}
where $\alpha_{X} = g_X^2/4\pi$. 
From an experimental bound (23) one can find that 
\begin{equation}
\alpha_{X} < 10^{-8}
\end{equation}
Light $X$-boson leads to additional contribution 
$\alpha_{X}/2\pi$ to 
the anomalous magnetic moment of electron. From the bound (19) 
we  find that 
\begin{equation}
\alpha_{X} \lesssim 3 \times 10^{-10},
\end{equation}
\begin{equation}
Br(X \rightarrow \gamma \gamma X) \lesssim 1.2\cdot 10^{-7}
\end{equation}
The bound (28) is approximately 30 times  smaller than the 
experimental bound (23). Note 
that if $X$-boson 
interacts only with the right-handed or left-handed electrons 
such interaction does not give contribution at one loop level 
to $a_e$  and the 
contribution to the anomalous magnetic moment of the electron is 
of the order $O(\alpha \alpha_{X}/\pi^2)$. 
As a consequence we find 
that the anomalous magnetic moment of electron gives much weaker bound 
$\alpha_{X} \lesssim O(10^{-7})$.

\subsection{Exotic $o-Ps\to 2\gamma$ and $o-Ps\to 4 \gamma$ decays}

The decay $o-Ps \rightarrow 4 \gamma $ is 
forbidden in QED by C-invariance. The corresponding experimental bound 
is \cite{yang}
\begin{equation}
Br(o-Ps \rightarrow 4 \gamma) <2.6 \times 10^{-6}
\end{equation}
The $o-Ps$ has spin one and cannot decay into two photons due to conservation 
of angular momentum which  follows from 
the isotropy of space. Hence, searching for decays   
$o-Ps \rightarrow 2 \gamma $ tests spatial isotropy. 
The best current result is \cite{asa}
\begin{equation}
Br(o-Ps \rightarrow 2 \gamma) < 3.5 \times 10^{-6} 
\end{equation}

\section{Invisible decays of orthopositronium}

The experimental signature of  the $o-Ps\rightarrow invisible$ decay  
is the absence of an energy deposition of $\sim$ 1 MeV,
which is expected from the ordinary $o-Ps$ annihilation,
in a 4$\pi$ hermetic calorimeter surrounding the $o-Ps$ formation region. 
The first experiment on $o-Ps\to invisible$ decay was performed  
a long time ago \cite{atojan}, and then was repeated 
with a higher sensitivity  by Mitsui et al. \cite{mits}. 
The best current experimental bound is \cite{mits} 
\begin{equation}
Br(o-Ps \rightarrow invisible ) < 2.8 \times 10^{-6} 
\end{equation}

Consider several motivations for further experimental searching 
for this decay mode.\\

\subsection{Millicharged particles}

The first one is related to the fundamental problem of charge 
quantization. 
In 1986, Holdom \cite{holdom} showed that, by adding a second, 
unobserved, photon (the ``shadow'' photon) one could construct in a 
natural way grand unified models, which contain particles with an electric 
charge very small compared to the electron charge.
This work has stimulated new theoretical and experimental tests
(for a recent review see \cite{davidson} and references therein).
If such  milli-charged particles exist with a small mass, 
the $o-Ps$ could decay apparently invisibly, since the particles 
would mostly penetrate any type of calorimeter  without interaction.
 The corresponding decay width is \cite{ignatiev}
\begin{equation}
\Gamma(o-Ps \rightarrow X \bar{X}) = \frac{\alpha^5 Q^2_{X}m_e}{6}\cdot 
k \cdot F(\frac{m^2_X}{m^2_e}),
\end{equation}
where $Q_{X}$ is an electric charge of the $X$-particle $(Q_e \equiv 1)$,
 $k = 1$, 
$F(x) = (1 - \frac{1}{2} x)(1-x)^{\frac{1}{2}}$ for spin $\frac{1}{2}$ and
  $k = \frac{1}{4}$,
$F(x) = (1-x)^{\frac{3}{2}}$ for spin-less $X$-particle. 
For spin $\frac{1}{2}$ millicharged $X$-particle and for $m_X \ll m_e$ one 
can find from experimental bound  (31) 
that  $Q_{X} \lesssim 8.6 \cdot 10^{-5}$  \cite{mits}. Search 
for the $o-Ps \to invisible$ decay with the sensitivity 
$Br(o-Ps \to invisible)\simeq 10^{-8}$ would touch  the parameter space 
not excluded by the resuls of the recent experiment at SLAC \cite{prinz}. 

\subsection{Mirror World}

In refs. \cite{glashow} - \cite{foot} an explanation of 
orthopositronium decay rate puzzle  within  
the model with the Mirror Universe \cite{mirror} 
has been discussed.\footnote{For many interesting ideas on mirror matter 
see e.g. ref. \cite{foot1} and  
recent book \cite{foot2} and references therein.} The existence of 
the effective  mixing $\epsilon F^{\mu\nu}F^{'}_{\mu\nu}$ between 
our photon and mirror photon would  mix ordinary 
and mirror orthopositronium, see e.g. \cite{ignasha}, resulting in
$o-Ps$ - mirror $o-Ps$ oscillations \cite{glashow}. 
This leads to an apparent increase 
in the $o-Ps$ decay rate. Since the mirror decays are not detected,
the experimental signature of this effect is $o-Ps\to invisible$ decay with 
the branching ratio
\begin{equation}
Br(o-Ps \rightarrow invisible) =
\frac{2(2\pi\epsilon f)^2}{\Gamma^2 + 4 (2\pi\epsilon f)^2},
\end{equation}
where $\epsilon, \Gamma, $ and $f$ are, respectively the mixing 
strength between the photon and mirror photon, the decay rate of $o-Ps$ 
into three photons and the contribution to the 
ortho-para splitting from the one-photon annihilation diagram involving 
$o-Ps$ ($f = 8.7 \times 10^4~MHz$) \cite{glashow}. 
The limit on photon-mirror photon mixing strength  extracted from the
bound (31) taking into account the suppression collision factor,
 is $\epsilon < 10^{-6}$ \cite{gnin} and it is
not strong enough to exclude a possible mirror contribution  to
the $o-Ps$ decay rate. 
A vacuum experiment on $o-Ps\to invisible$ decay 
with the sensitivity to the mixing strength  $\epsilon \simeq 10^{-7}$ is
necessary to confirm or rule out the mirror world effect \cite{foot}. 

\subsection{New light $X$-boson}

Consider now  the model with a light vector $X$-boson and with 
the interaction  of Eq.(24) which  also leads to invisible decay of 
$o-Ps$.\footnote{For the recent  phenomenological bounds in models with 
light vector $X$-boson related to the muon $(g-2)$ and, 
so-called NuTeV anomalies see, respectively \cite{gnkr} and \cite{david}, 
and also \cite{dob}.} Suppose in addition $X$-boson interacts with 
other particles (fermions) or (as a consequence of the Higgs mechanism) 
with itself and scalar field.
The contribution of the $X$-boson to the electron anomalous 
magnetic moment is given by the well known formula
\begin{equation}
\delta a_e = \frac{\alpha_{X}}{\pi}\int^{1}_{0}\frac{x^2(1-x)}{x^2 +(1-x)
\frac{m^2_X}{m^2_e}}
\end{equation}
For $m_X \ll m_e$ from the bound (19) we find that  
$\alpha_{X} < 3 \times 10^{-10}$. For the opposite case 
of heavy $X$-boson ($m_X \gg m_e$) the bound on the anomalous 
electron magnetic moment leads to the bound
\begin{equation}
\alpha_{X}\frac{m^2_e}{m^2_X} < 4.5 \cdot 10^{-10}
\end{equation}
For the reaction $o-Ps \rightarrow X^{*} \rightarrow X_1 \bar{X}_1$ 
(here $X^{*}$ is virtual $X$-boson and $X_1$ is a fermion 
(sterile neutrino) or a scalar particle) one can find 
that
\begin{eqnarray}
Br(o-Ps \rightarrow X^{*} \rightarrow X_{1}\bar{X}_{1}) 
= \\ \nonumber
\frac{3\pi}{4(\pi^2 - 9)}
\cdot k \cdot F(\frac{m^2_{X_1}}{m^2_e})   
(1 -\frac{m^2_{X}}{m^2_{0-Ps}})^{-2}\frac{\alpha_{X} \alpha_{XX_1}}
{\alpha^3},
\end{eqnarray}
where $F(x)$ has been defined before and 
$\alpha_{XX_1} = g^2_{XX_1}/4\pi$.  From the bound (19)  we find 
for $m_X \ll m_e$ that 
\begin{equation}   
 Br(o-Ps \rightarrow X^{*} \rightarrow X_{1}\bar{X}_{1}) \lesssim
k \times 2 \cdot 10^{-3}\cdot \alpha_{XX_1}
\end{equation}
To have experimentally interesting branching of the order 
$O(10^{-6})$, setting  $\alpha_{X} = 3 \times 10^{-10}$ we must 
have $\alpha_{XX_1} \sim
5 \cdot 10^{-4}$.  

In the opposite limit $m_X \gg m_e$ the corresponding bound reads
\begin{equation}   
 Br(o-Ps \rightarrow X^{*} \rightarrow X_{1}\bar{X}_{1}) \leq
k \times 3 \cdot 10^{-3}\cdot \alpha_{XX_1} \cdot \frac{m^2_e}{m^2_X}
\end{equation}
For $X$-boson mass close to 
the orthopositronium mass, we have the enhancement factor 
$(1 - \frac{m^2_X}{m^2_{o-Ps}})^{-2}$ in formula (36) for the branching 
 ratio and hence the coupling constant $\alpha_{X}\alpha_{XX_1}$ could be 
smaller.

 The discussed above explanation \cite{glashow} of the o-Ps 
lifetime discrepancy based on the model with mirror  Universe 
 \cite{glashow} in fact uses the enhancement factor 
$(1 - \frac{m^2_X}{m^2_{o-Ps}})^{-2}$. According to our terminology 
we can treat $X$-boson as 
a mirror orthopositronium which due to mixing of our photon 
with mirror photon has direct coupling with electrons and decays into 
3 invisible mirror photons.   
Vector boson $X$ acquires a mass via Higgs mechanism
and for both light $X$-boson and the  Higgs scalar 
$\phi$ the reaction
(``Higgs-Strahlung'' process) 
\begin{equation}
o-Ps \rightarrow X^{*} \rightarrow X \phi
\end{equation}
also leads to invisible $o-Ps$ decay mode. 
The corresponding formula for the $o-Ps$ branching in the limit 
$m_x \ll m_e$, $m_{\phi} \ll m_e$ coincides with formula 
(36) for $k = \frac{1}{4}$ (scalar case) and leads to similar bound 
on the product $\alpha_{X}\alpha_{X\phi}$. 

\subsection{Extra dimentions}

Recently  the models with infinite 
additional dimensions  
of Randall-Sundrum type (brane-world models) 
\cite{randall},\cite{rubakov} have become very popular. 
There is a  hope 
that  models with a big compactification radius \cite{randall}-\cite{giud}
 will provide the natural solution to the 
gauge hierarchy problem.
For instance, as it has been shown \cite{randall} 
in the five dimensional model,  there exists a 
thin-brane solution to the 5-dimensional Einstein equations which 
has flat 4-dimensional hypersurfaces,
\begin{equation}
ds^2 = a^2(z)\eta_{\mu\nu}dx^{\mu}dx^{\nu} -dz^2.
\end{equation} 
Here
\begin{equation}
a(z) = exp(-k(z-z_c))
\end{equation}
and the parameter $k > 0 $ is determined by the 5-dimensional Planck 
mass and bulk cosmological constant. For the model with 
metric (40) the effective four-dimensional gravitational constant is
\begin{equation}
G_{(4)} = G_{(5)}k \frac{1}{exp(2kz_c) - 1}
\end{equation}
One can solve the gauge hierarchy problem in this model if  
$k \sim M_{EW} =  1~TeV$, $G_{(5)}\sim k^{-3}$. As it follows 
from the expression (42) the Planck scale in this model is
\begin{equation}
M_{PL} \sim exp(kz_c)M_{EW}
\end{equation}
that means the existence of exponential hierarchy between Planck 
and electroweak scales. For $z_c \approx 37 \cdot k^{-1}$ we have 
correct quantitative relation among Planck and electroweak 
scales. Note that in this model 
the mass of the first gravitational Kaluza-Klein state is 
$m_{grav} \sim k$. 
As it has been shown in ref.\cite{tinyakov} 
 the massive matter becomes 
unstable due to tunnelling effect
 and disappears into additional 5-th dimension. For 
massive scalar particle $\Phi$ with the mass $m$ the transition rate 
into additional dimension is given by the 
formula \cite{tinyakov}
\begin{equation}
\Gamma(\Phi \rightarrow additional~dimension) =
\frac{\pi m}{16}(\frac{m}{k})^2
\end{equation}
 For  massive vector 
particles the expression for its transition rate into additional 
dimension(s) is  not explicitly known. 
 To make quantitative estimate we 
shall use the formula (44) for the decay width of 
spin 1 particle. 
      
We stress that $o-Ps$ is a good candidate 
for the searching for effect of disappearance into 
additional dimension(s) since it has specific quantum numbers similar 
to those of vacuum and is a system which allows its constituents a rather 
long interaction time.
For the orthopositronium invisible decay into 
additional dimension(s) 
\begin{equation}
o-Ps \rightarrow \gamma^{*} \rightarrow  additional~dimension(s) 
\end{equation}
the corresponding branching ratio is
\begin{eqnarray}
Br(o-Ps \rightarrow \gamma^{*} \rightarrow additional~dimension(s))& 
=\\ \nonumber 
\frac{9\pi}{4(\pi^2-9)}\cdot \frac{1}{\alpha^2}\cdot \frac{\pi}{16}
(\frac{m_{o-Ps}}{k})^2   \approx 3\cdot10^{4}(\frac{m_{o-Ps}}{k})^2
\end{eqnarray}
Important bound on the parameter $k$ arises from data on $Z\to invisible$ 
decay. LEP1 bound 
 on the number of neutrinos $n_{\nu} = 3.00 \pm 0.06$ \cite{particle} 
extracted from direct measurements of
the  $\Gamma(Z \to invisible)$ decay rate 
results  at the $2\sigma$-level in the bound 
$\Gamma(Z \rightarrow additional~dimension(s)) \lesssim 0.02~GeV$ 
and leads to  $k  \gtrsim 2.7~TeV$. Using this we find  
\begin{equation}
Br(o-Ps \rightarrow additional~dimension(s)) \lesssim 4 \cdot 10^{-9}
\end{equation}
To solve the gauge hierarchy 
problem models with additional infinite dimension(s) 
must have the $k \lesssim O(10)~TeV$. It means that   
\begin{equation}
Br(o-Ps \rightarrow additional~dimension(s)) \gtrsim O(10^{-10})
\end{equation}
Since these estimates give only an order of magnitude for the 
lower and upper limits on corresponding branching ratio, we believe
that the region of $Br(o-Ps\to invisible) \simeq 10^{-9}-10^{-8}$ is 
of great interest for observation of effect of extra dimensions.
However, more accurate calculations of the tunnelling effect
might provide more stringent bounds on transition rate of 
$o-Ps$ into extra dimension(s).


\section{Conclusion}

Due to its specific properties, orthopositronium is
 an important probe of QED and also physics beyond the Standard Model.
We have reviewed phenomenological models of 
rare $o-Ps$ decay modes and updated some existing bounds. 
 We have considered a model with infinite additional dimension(s) in which
orthopositronium may disappear as a result of tunnelling into additional 
infinite dimension(s).
The experimental signature of this effect is the  invisible decay
of orthopositronium which may occur at a rate within several
 (two or three) orders of magnitude of the present experimental upper limit.
This result strengthens  current motivations and justify efforts for a 
more sensitive search of the $o-Ps\to invisible$ decay 
in a near future experiment \cite{ethz}. 

N.V.K. is indebted to V.A.Rubakov for many useful discussions on 
models with additional infinite dimensions.

\end{document}